\documentclass{aa}
\usepackage{graphicx,epsfig}
\usepackage{amssymb,amsmath}

\newcommand{\etal}{{et al.\ }}
\newcommand{\lta}{\lesssim}
\newcommand{\gta}{\gtrsim}

\begin{document}

\title{The Impact of Starbursts and Post-Starbursts on the Photometric 
Evolution of High Redshift Galaxies} 

\author{Uta Fritze -- v. Alvensleben$^1$, Jens Bicker$^2$}
\institute{$^1$ Centre for Astrophysics Research, 
University of Hertfordshire, Hatfield, Herts AL10 9AB, UK \\
$^2$ Institut f\"ur Astrophysik, Universit\"at G\"ottingen, Friedrich-Hund-Platz 1, 
37077 G\"ottingen, Germany}
\authorrunning{Fritze -- v. Alvensleben \& Bicker}
\titlerunning{Starbursts and postbursts at high redshift}
\offprints{U. Fritze -- v. A., \email{ufritze@star.herts.ac.uk}}
\date{Received xxx  / Accepted xxx}

\abstract{We present evolutionary synthesis models for galaxies of 
spectral types Sa through Sd 
with starbursts of various strengths triggered at various redshifts and 
study their
photometric evolution before, during, and after their bursts in a cosmological
context. We find that bursts at high redshift, even very strong ones, 
only cause a small 
blueing of their intrinsically blue young parent galaxies. At lower redshift, 
in contrast, even small bursts cause a significant blueing of their 
intrinsically redder 
galaxies. While the burst phase is generally short, typically a few hundred 
Myr in normal-mass galaxies,
the postburst stage with its red colors and, in particular the very red ones 
for early bursts at high
redshift, lasts much longer, of the order of several Gyr. We find that,  
even without any dust, which in the postburst stage is not expected to play 
an important role
anyway, models easily reach the colors of EROs in the redshift range z 
$\sim 2$ through z $\sim 0.5$ after starbursts at redshifts between 2 and
4. We therefore propose a third alternative for the ERO galaxies beyond the two 
established ones of passive galaxies vs. dusty starbursts: the
dust-free post-(strong-)starbursts. 
A very first comparison of our models to HDF data with photometric redshifts 
shows that
almost all of the outliers that could not be described with our chemically
consistent models for undisturbed normal 
galaxy types E through Sd can now be explained very well. 
Galaxies in the redshift
range from z$\sim 2.5$ to z$\sim 0.5$ that are redder, and in some cases much
redder, than our reddest undisturbed model for a high-metallicity classical
elliptical are well described by post-starburst 
models after starbursts at redshifts between 2 and
4. Galaxies bluer than our bluest low metallicity Sd model, most of which have 
redshifts lower than 1, are well explained by ongoing starbursts.}

\maketitle

  \keywords{Galaxies: formation \& evolution, starbursts, high-redshift}


\section{Introduction}
\label{INTRO}
\subsection{Importance of starbursts at high redshifts}
Despite considerable effort and advances from both the theoretical and
observational sides, some very fundamental questions about the formation and
evolution of galaxies are still open. One of them concerns the respective roles
of hierarchical merging versus accretion in the evolutionary history of
galaxies and their respective impacts on the star formation (SF) histories and
on the final -- or better today's -- morphological and spectral appearance of
galaxies. Major
mergers, i.e. mergers involving two objects of comparable mass, clearly
trigger strong starbursts, as long as they involve sufficient quantities of
gas, as shown by examples in the local universe, like NGC 7252 (Schweizer
1982, Fritze -- v. Alvensleben \& Gerhard 1994a \& b) or the Luminous IR Galaxies
(=LIRGs), like the Antennae (Whitmore \& Schweizer 1995), with their global starbursts, or, even
stronger, the Ultra Luminous IR Galaxies (ULIRGs), all of which feature
extremely strong nuclear starbursts, sometimes around an AGN, and are in
advanced stages of merging (cf. Genzel \etal 2001). 

Minor mergers with mass ratios $\gta 3:1$ under the same condition can trigger
smaller starbursts. The situation is less clear for accretion. Slow and steady
accretion need not trigger a starburst directlybut may do so in an indirect way
by slowly destabilizing a disk.  Accretion of wholesale lumps of gas (and
stars) up to dwarf galaxies, on the other hand, may enhance SF significantly.
The boundary, however, between enhanced SF and what would be called a burst is not very
precisely defined. Classically, a starburst is defined as a rate of SF that is
too high to be sustained for a long time without exhausting a galaxy's gas
reservoir. With infall, this definition gets somewhat weakened.

Even within the framework of a given structure and galaxy formation scenario,
like the hierarchical merging picture, at one extreme, and the initial collapse
plus successive accretion scenario -- e.g. from filaments, as proposed by 
Bournaud \etal (2005) -- on the other, it is not clear how much
happens at which redshift. Recent detections of unexpectedly massive, evolved, 
and numerous galaxies at high redshift (Chapman \etal 2004, Cimatti \etal 2004,
Franx \etal 2003, Le Floc'h \etal 2004) have challenged at least the
semi-analytic version of the otherwise widely accepted $\Lambda-$CDM models. 

Many different and independent pieces of evidence indicate that the merger rate
was higher in the past. How much higher, however, is still controversial (e.g.
Zepf \& Koo 89, Wu \& Keel 1998, Conselice 2003). 

Galaxies were richer in gas in the past, so that mergers should have caused even
stronger starbursts than locally. Damped Lyman-$\alpha$ systems, abundant in
number density at ${\rm z \sim 2 - 4}$ and containing the bulk of the baryonic
matter at these redshifts in the form of HI (e.g. Wolfe \etal 1993), are
believed to be progenitors of present-day spirals and to already contain
tremendous amounts of mass, almost all of it still in the form of gas
(Prochaska \& Wolfe 1998). If those objects were to be involved in a merger, the
reservoir for a starburst would be gigantic.  Indeed, there is even direct
evidence that starbursts are stronger towards higher redshifts. The space
densities of LIRGs and ULIRGs increase with redshift, and new classes of
extremely powerful starbursts appear: hyper-luminous IR galaxies and SCUBA
sources. Their dust-enshrouded SF rates are estimated to be of the order of 
thousand(s) of solar masses per year (e.g. Chapman \etal 2004, Le Floc'h \etal
2004). 

Many Lyman Break Galaxies (LBGs) are also called starbursts by some authors,
albeit with more moderate SF rates than ULIRGs or SCUBA galaxies, typically in
the range ${\rm 10 - 100~M_{\odot}/yr}$ (Shapley \etal 2001, Pettini \etal
2001, Giavalisco 2002). Although some of them feature irregular or knotty
structures, they are not clearly related to (major) mergers. It should be kept
in mind that even the good old monolithic initial collapse scenario
(Eggen, Lynden -- Bell \& Sandage 1962), as well as many different pieces of
evidence -- from the present stellar masses through chemical
enrichment to fossil records -- predict, that the SF rates of normal Hubble
sequence galaxies should have been higher in the past (e.g. Sandage 1986). 
Hence, the discrimination between ``normal SF`` and a burst also gets 
less clear-cut towards larger look-back times. 

All in all, the currently emerging scenario of global cosmic SF describes two
phases: at high redshift SF in galaxies was dominated by
discrete, recurrent bursts, possibly associated with mergers. Around ${\rm z
\sim 1.4}$ the universe switched gears and SF in galaxies became more
quiescent, governed by accretion more than by mergers, allowing the Hubble
sequence to emerge (cf.  Papovich \etal 2005). With respect to individual
galaxies, both the mass assembly and the bulk of SF appear to have occurred
earlier in galaxies that by today are very massive than in lower mass galaxies
where the bulk of present-day SF is still taking place.  Extending results by
Heavens \etal (2004) on the fossil record of SF in $\sim 10^5$ nearby galaxies
from the SDSS, Hammer \etal (2005) present additional evidence from a sample of
CFRS galaxies at ${\rm z>0.4}$ for a scenario in which intermediate mass
galaxies ${\rm (3-30) \times 10^{10}~M_{\odot}}$ formed a large fraction of
their stars only 4 to 8 Gyr ago, most probably in a series of recurrent bursts
of SF, many of which apparently gave rise to LIRG phases. The high fraction of
LIRGs among intermediate mass galaxies at ${\rm z>0.4}$, together with a
fraction of $> 17$\% of those LIRGs showing evidence for major mergers, supports
this hypothesis. 

All this clearly shows the important role of starbursts (and mergers) at high
redshift. 

\subsection{Motivation for the study of starburst and post-starburst models}
A comparison between our set of undisturbed, chemically consistent galaxy 
models E
through Sd over the redshift range from ${\rm z>4}$ through ${\rm z \gta 0}$ 
and the 
Hubble Deep Field North (HDF-N) galaxies with photometric redshifts from 
Sawicki et al. (1997) was presented in Bicker et al. (2004). We recall 
that the agreement between our chemically
consistent models, which account for the increasing metallicities of successive
stellar generations, and the data was much better than for any earlier models
using solar metallicity input physics alone. The inclusion of stellar
subpopulations with lower-than-solar metallicity made the chemically
consistent Sd model bluer at increasing redshifts than an equivalent model with
solar metallicity alone. On the other hand, including stars with 
supersolar metallicity made our chemically consistent, classical elliptical 
model redder at intermediate redshifts than the corresponding model for ${\rm
Z_{\odot}}$. Nevertheless, while we found our chemically
consistent galaxy models to bracket the redshift evolution of the bulk of 
the HDF data well over the redshift range 
from ${\rm z>4}$ through ${\rm z \gta 0}$ between the red elliptical and the
blue Sd models, there remained a number of galaxies
with bluer colors than those of our bluest undisturbed Sd model, described by a
constant SFR and fairly low metallicity, at low and intermediate redshifts, as
well as quite a few galaxies with significantly redder colors than for our reddest
undisturbed elliptical galaxy model -- despite its high metallicity in the 
redshift range ${\rm 0.5 \lta z \lta 4}$. This motivated us
to investigate if and to what extent starbursts can explain 
the ``bluer-than-normal'' galaxies and if and to what extent 
post-starbursts can explain the 
``redder-than-normal'' ones. 

\subsection{The present study}
We here present models for galaxies with starbursts of various strengths 
occurring at various evolutionary stages or redshifts to study the impact of
ongoing and past starbursts on the photometric properties of these galaxies.   
While spectra, of course, can give fairly precise SF rates, 
the advantage of the photometric signatures of starbursts and post-starbursts is
that they can be compared to the wealth of multi-wavelength imaging data
available for several deep fields, and help constrain the rates, strengths, and fate
of starbursts for huge numbers of galaxies, most of which are too
faint for spectroscopy. 

The widely used SED (spectral energy distribution) fitting technique to obtain 
photometric redshifts and galaxy types -- using observed or model SEDs -- gives 
the number of ongoing starbursts as a function of redshift. Only the 
evolutionary synthesis approach presented here, however, allows a
galaxy to be followed through a starburst and beyond in a cosmological context and, hence, 
to identify the successors of high-redshift starbursts in terms of 
post-starburst galaxies at lower redshifts.  

The outline of the present paper is the following. In Sect. 2.1, we
briefly present our evolutionary synthesis code GALEV that has already been
extensively described in the context of evolution models for star clusters and
``normal'' undisturbed galaxies, including cosmological and evolutionary
corrections from high to low redshift. We describe how we
put starbursts on top of our undisturbed galaxy models and discuss the grid of
parameters we chose to study in Sect. 2.2 and 2.3.  In
Sect. 3.1, we present selected results and show the impact of bursts
of various strength at various times/redshifts on the luminosities and colors
of galaxies from the optical to NIR, not only during the active burst phase but
also after the bursts. In Sect. 3.2 and 3.3, we discuss different burst
strengths and the role of dust. In
Sect. 3.4, we investigate in how far our models relate to the Extremely
Red Objects (EROs) in their postburst phases and propose a new scenarion for the
passively evolving 50 \% of the ERO population in terms of (dust-free)
post-strong-starbursts. Finally, Sect. 3.5 shows a first comparison
with Hubble Deep Field (HDF) data. A consistent interpretation of HDF 
galaxies in terms of our chemically consistent models for undisturbed galaxies, the
starburst and postburst models as presented here, our own photometric redshifts 
and a detailed quantitative analysis in 
terms of burst strengths, starburst and post-starburst galaxy fractions, 
metallicity and mass evolution, etc., will be the subject
of a forthcoming paper.

\section{Models}
\label{MODELS}
\subsection{Undisturbed galaxies}
\label{UDGAL}
Our evolutionary synthesis code GALEV in its latest version has been described
in the context of the chemically consistent evolution of galaxies (Bicker et
al. 2004).  The code accounts for the increasing initial metallicities of
successive stellar generations and naturally reproduces the observed broad
stellar metallicity distributions of local galaxies. 
It is based on isochrones for 5 different
metallicities ${\rm -1.7 \leq [Fe/H] \leq +0.4}$ from the Padova group that
include the thermal-pulsing AGB phase, which has been shown by Schulz \etal 
(2002) to have an important
effect on colors like ${\rm (V-I)}$ and ${\rm (V-K)}$ at ages from 100 Myr 
through a few
Gyr for single-burst, single-metallicity stellar populations like star 
clusters.

The spectral evolution of a stellar population is calculated on the basis of
the Lejeune \etal (1997, 1998) library of model atmosphere spectra for the
five metallicities.  Luminosities in a wide variety of filter
systems and colors are obtained by folding the spectra with the respective
filter response functions. In Anders \& Fritze -- v. Alvensleben (2002) we
added the emission contributions from gas ionized by young stars both in terms
of continuous and line emission for the respective metallicities to our single
burst single metallicity models, while in Bicker \& Fritze -- v. Alvensleben (2005) we
discussed its effects in the context of galaxies. 

The various spectral types of undisturbed galaxies  from E through Sd are
described by their respective appropriate SF histories.  The spiral models Sa,
Sb, Sc, Sd discussed in the present context use SF rates linearly tied to the
evolution of the gas content that decreases through SF and increases through
stellar mass loss. For the model of a classical initial-collapse elliptical
galaxy, we use a SFR exponentially declining with time with an e-folding time 
of 1 Gyr. Constants
are chosen so as to match, after $\sim$13 Gyr, the observed average
colors, gas content, characteristic HII region metallicities, M/L-ratios, and
template spectra of the respective spectral types in the local universe. Models
do not include any spatial resolution, dynamics, or AGN contributions. 
No infall is included,
galaxies are described by closed box models for simplicity. We have shown that
continuous infall that increases the mass of a galaxy by no more than a
factor of $\sim 2$ does not significantly affect the time evolution of 
spectra, colors, 
and metallicities, since SF histories in this case need to be adjusted
in order to still match the present-day observables. To a large extent, this 
cancels the effects of this type of infall on the color and metallicity
evolution (Fritze -- v. Alvensleben 2000).

A standard cosmological model with ${\rm H_0=70,~\Omega_m=0.3,
~\Omega_{\Lambda}=0.7}$, and a redshift of galaxy formation of ${\rm
z_f \sim 5}$ are assumed to calculate redshifted galaxy spectra, luminosities, 
and colors. For galaxies at high redshifts the average attenuation by
intergalactic HI along the line of sight as determined by Madau (1995) is
included in the models (cf. M\"oller et al. 2001). Dust within the galaxies,
however, is not yet included at the present stage, neither in terms of absorption 
in the UV and optical nor in terms of thermal reemission in the mid- and 
far-IR.

Gaseous emission in terms of lines and continuum is consistently included in our
models for the respective metallicities (cf. Bicker \& Fritze - v. Alvensleben
2005). It has been shown to have a large impact, even on broad band
luminosities and colors, during strong starbursts by Kr\"uger et al. (1995) (see
also Anders \& Fritze - v. Alvensleben 2003). 

Our undisturbed model galaxies are
scaled to match the mean present day observed ${\rm M_B}$ of the respective 
galaxy types in the local universe as determined by Sandage et al. (1985). 
These average absolute 
B-band luminosities and their 
1-$\sigma$ ranges are ${\rm \langle M_B \rangle 
=-20.8\pm1.7}$, ${\rm \langle M_B \rangle =-18.1\pm0.94}$, and 
${\rm \langle M_B \rangle =-17.0\pm0.77}$ mag for E, Sb, and Sd galaxies, 
respectively (cf. Bicker et al. 2003 \& 2004).  

\subsection{Starbursts}
\label{BURST}
As a consequence of the assumption of ideal and instantaneous gas mixing 
and for the closed-box approximation used here, we cannot 
treat starbursts in a chemically consistent way. We therefore chose to 
model the bursting galaxies with fixed half-solar metallicity input physics   
in fair agreement with the high redshift galaxies with 
spectroscopic abundance determinations (e.g. Mehlert et al. 2002, Shapley et al.
2005). For consistency, the undisturbed 
galaxies used for comparison in this paper
are also calculated with fixed half solar metallicity. 

The bursts are described by a sudden increase in the SFR at the onset of a
burst at redshift ${\rm z_{burst}}$ or time ${\rm t_{burst}}$ to a value 
${\rm \Psi_{max}}$, followed by an exponential decline with an e-folding time 
or burst duration ${\rm \tau_{burst}}$ (cf. Fig. 1). 

\begin{figure}
  	\label{sfh}
	\centering{ 
	\epsfig{file=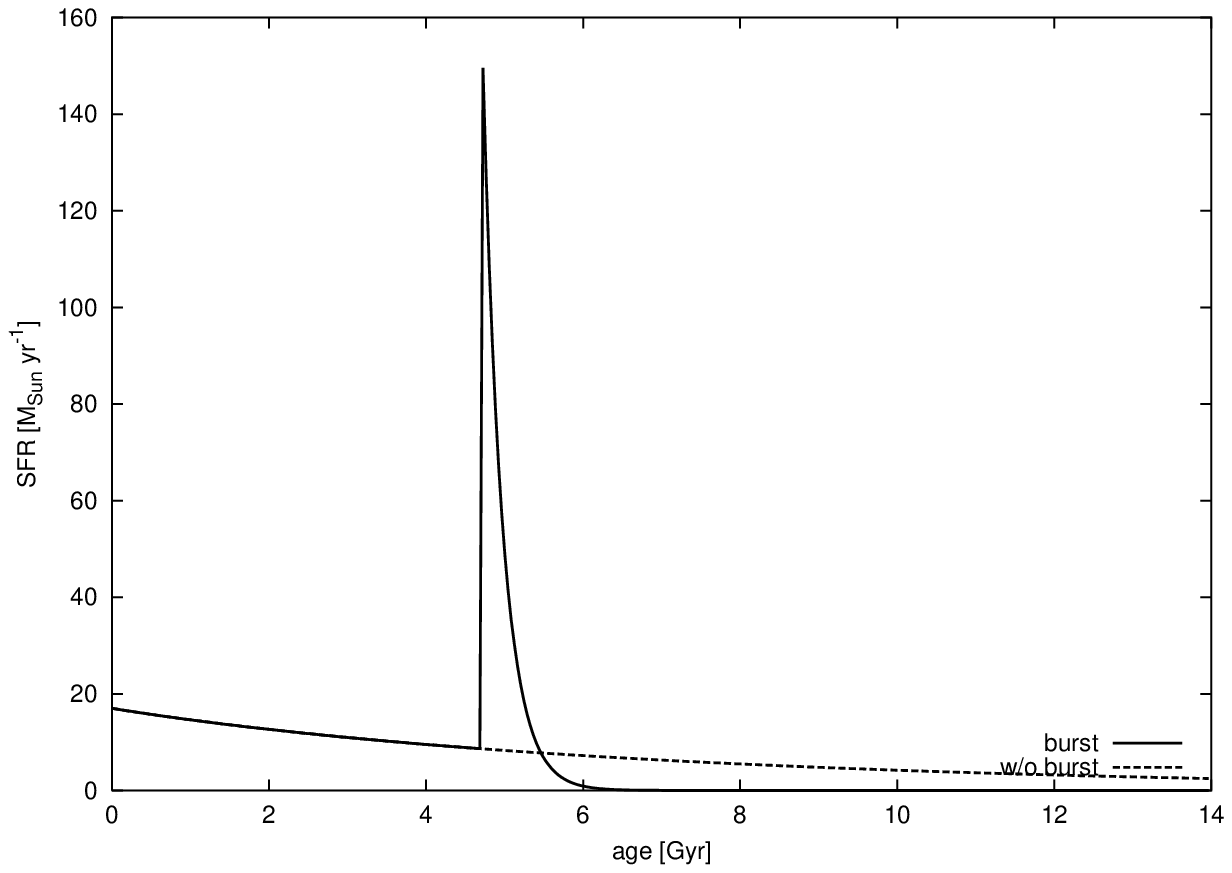,width=8.cm,angle=0}
       	\epsfig{file=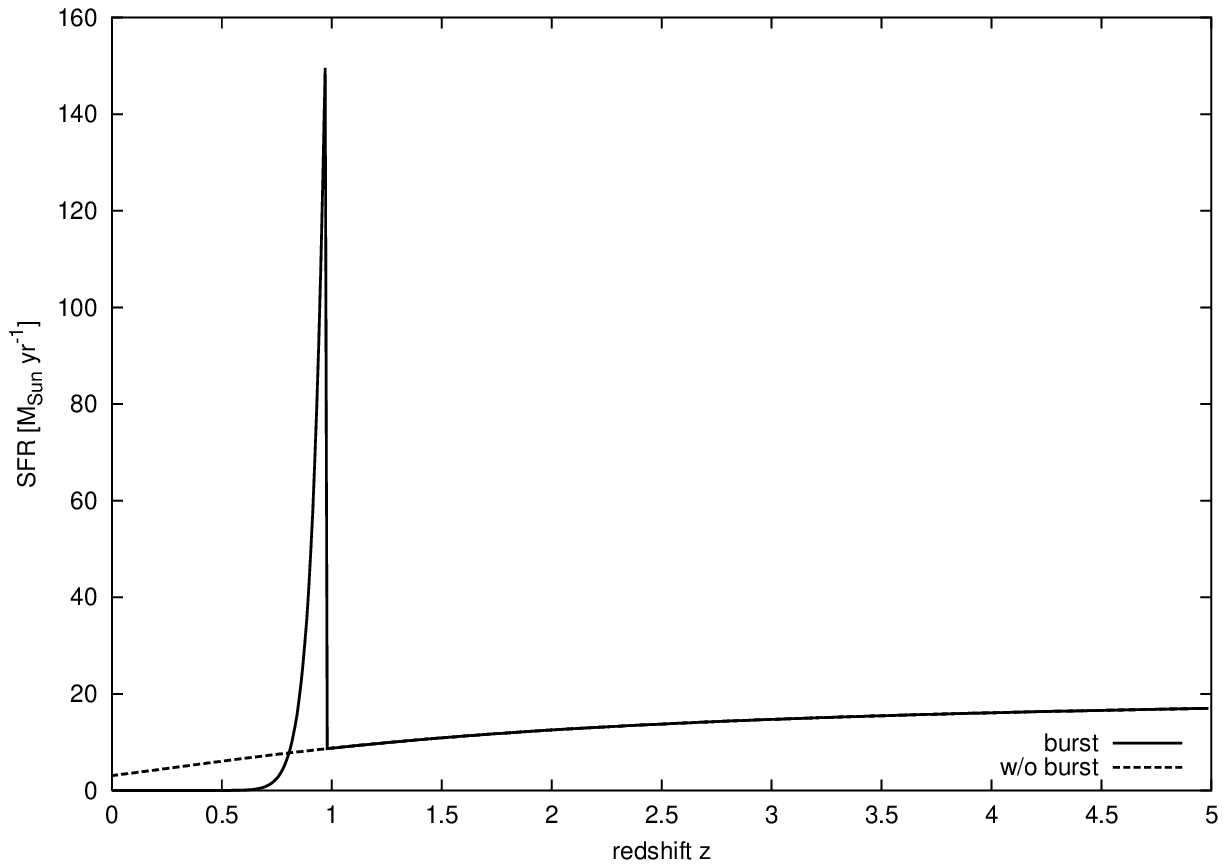,width=8.cm,angle=0}
  	}
\caption{Example of the SFH for a bursting galaxy. In this case a strong 
burst starts in an Sb galaxy at an age of 4.7 Gyr, corresponding to a 
redshift z $= 1$. Top panel: time evolution, bottom panel : 
redshift evolution.}
\end{figure}

This simple scenario is seen to cope well with observations of local starbursts
and post-bursts that become passive after the end of the
burst due to gas consumption by SF and, eventually, a loss of
left-over gas in a starburst-driven wind. We caution, however, that in the case of starbursts at high redshifts a
galaxy may regain some SF activity after a while, either due to accretion of 
primordial gas or due to re-accretion of gas that was formerly expelled. 

The strength of the burst b is defined by the fraction of gas available
at the onset of the burst ${\rm G(t_{burst})}$, which is
consumed for star formation during the burst (${\rm \Delta G}$):
\begin{eqnarray}
  {\rm b:=\frac{\Delta G}{G(t_{burst})}.}
\end{eqnarray}
Another common definition of the burst strength uses the fraction of
stars formed in the burst as compared to the total stellar mass of
the galaxy. We caution that, due to the evolving stellar-to-gaseous mass
ratio, there is no 1-to-1 relation between these two definitions. 

For clarity, we separately study starbursts within a given galaxy and 
starbursts induced by major ($=$ 1:1) mergers. All our figures refer to the 
first class of starbursts within a given
galaxy. A major merger between two galaxies of comparable type and 
luminosity leads to an additional increase in luminosity by 0.75 mag without 
affecting the colors. 

\subsection{Parameter ranges to be explored}
\label{PARAMETER}
Our undisturbed spiral galaxy models Sa, Sb, Sc, Sd are normalized to give 
by today, i.e at an evolutionary age of ${\rm t_0 \sim 13}$~Gyr, the respective mean local B-band
absolute luminosities observed by Sandage \etal
(1989) for the respective galaxy types. Bursts are assumed to be triggered at a range of redshifts ${\rm
z=4,~3,~2,~1,~0.5}$. Burst durations were shown by Fritze -- v. Alvensleben \&
Gerhard (1994a) to be a parameter that cannot be very tightly constrained. It is
to some degree degenerate with the detailed shape of the burst, i.e. steeply/slowly
rising/falling SFRs. Hence, we chose a fixed burst duration, i.e. $e$-folding
time for the exponential decrease in the SFR of ${\tau_b = 2.5 \times 10^8}$ yr,
as defined in Sect. 2.2. In general, burst durations appear to be related to
the dynamical timescales in galaxies. While dwarf galaxies feature burst
durations on the order of $10^6$ yr (Kr\"uger \etal 1991), normal galaxies -- the
objects of the present study -- are generally described by burst durations on 
the order of a few $10^8$ yr (e.g. Barnes \& Hernquist 1996, Fritze
-- v. Alvensleben \& Gerhard 1994a). 

The burst strengths we explore for this study range from a consumption of 20\%
to 50\% and 80\% of the available gas reservoir at the onset of the burst 
for weak, intermediate, 
and strong bursts, respectively. A general one-to-one transformation into  the
relative increases in stellar mass is not possible since the stellar mass and
the mass of the gas reservoir vary with redshift and galaxy type. 
For example, an intermediate strength (50\%) burst on top of an Sb type 
galaxy will increase
the stellar mass by a factor of 8.6 for a burst at redshift ${\rm z = 4}$, 
by a factor 2.3 for a burst at ${\rm z = 2}$, and by a factor 1.2 for a 
burst at ${\rm z = 0.5}$.

Note that at the present stage, our models do not include any dust, wether in
absorption or in its thermal reemission at rest-frame IR wavelengths. A 
consistent inclusion of appropriate amounts of dust in the course of the redshift
evolution of the various model galaxy types is difficult, in particular during
bursty phases of SF. Optical as well as FUV detected starbursts in the local
universe along with LBGs at high redshift, all do show dust extinction, 
albeit typically 
at rather moderate values of ${\rm A_{FUV}\sim 0.5 - 2}$ mag (Buat
\etal 2005, Shapley \etal 2001, Pettini \etal 2001), corresponding to visual 
extinctions on the order of ${\rm A_V \sim 0.2 - 0.9}$ mag. A comparison between FIR
(60 $\mu$m) and NUV (GALEX) selected starburst galaxy samples shows that all
but one out of 118 FIR-selected starbursts are also detected in the NUV and 
that the median NUV dust attenuation for the FIR-selected starbursts 
is only $\sim 2$ mag (Buat \etal 2005). Hence, while the present models 
clearly cannot be applied to
the interpretation of very dusty starbursts like those observed in ULIRGs or
SCUBA galaxies, they will certainly give a valid
approximation of the majority of more normal, i.e. not too dusty, starbursts. 

\section{Results}
\label{RESULT}
\subsection{Impact of starbursts on the photometric properties of galaxies}
\label{PHOTOM}
For our discussion of the model properties we use photometric data
in the standard Johnson-Cousins filters in the VEGAMAG system.

\begin{figure}
  \centering \epsfig{file=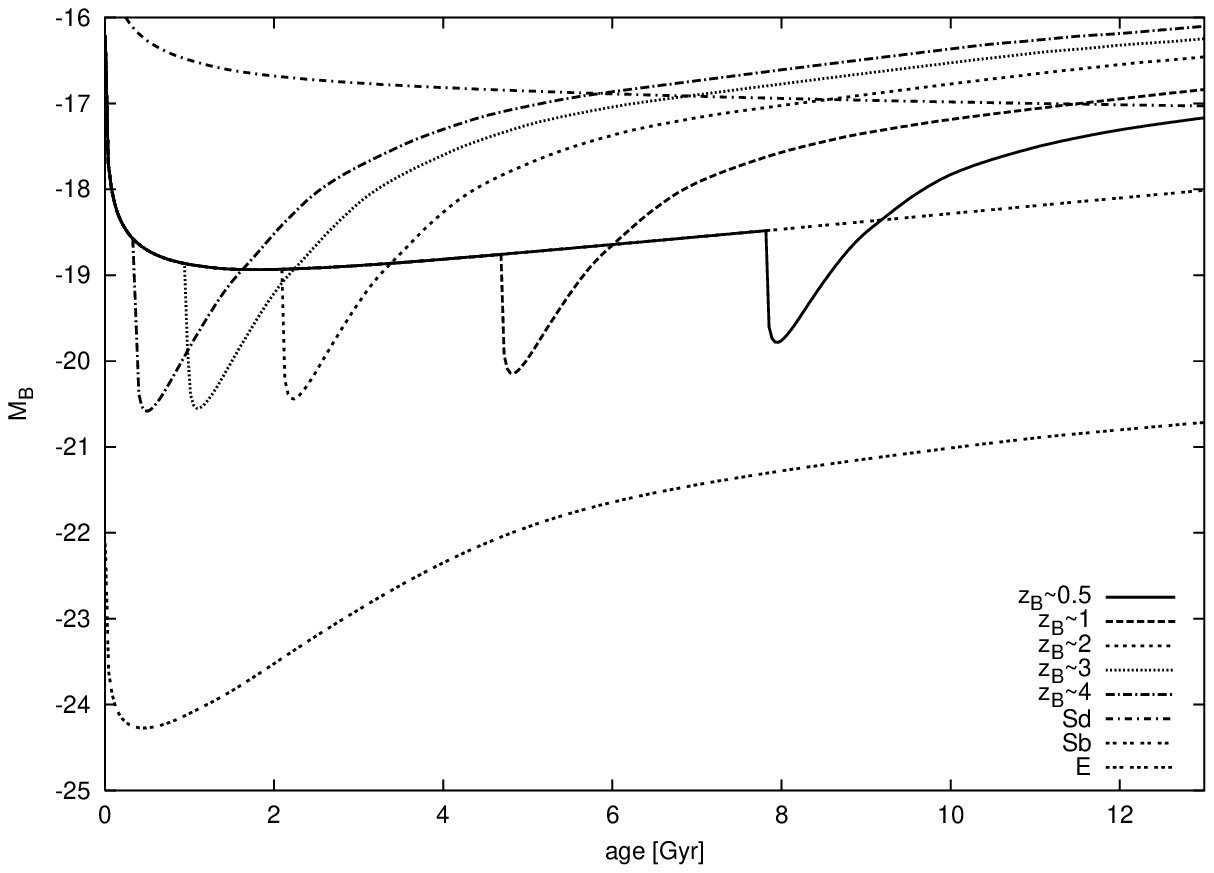,width=8.cm,angle=0}
  \centering \epsfig{file=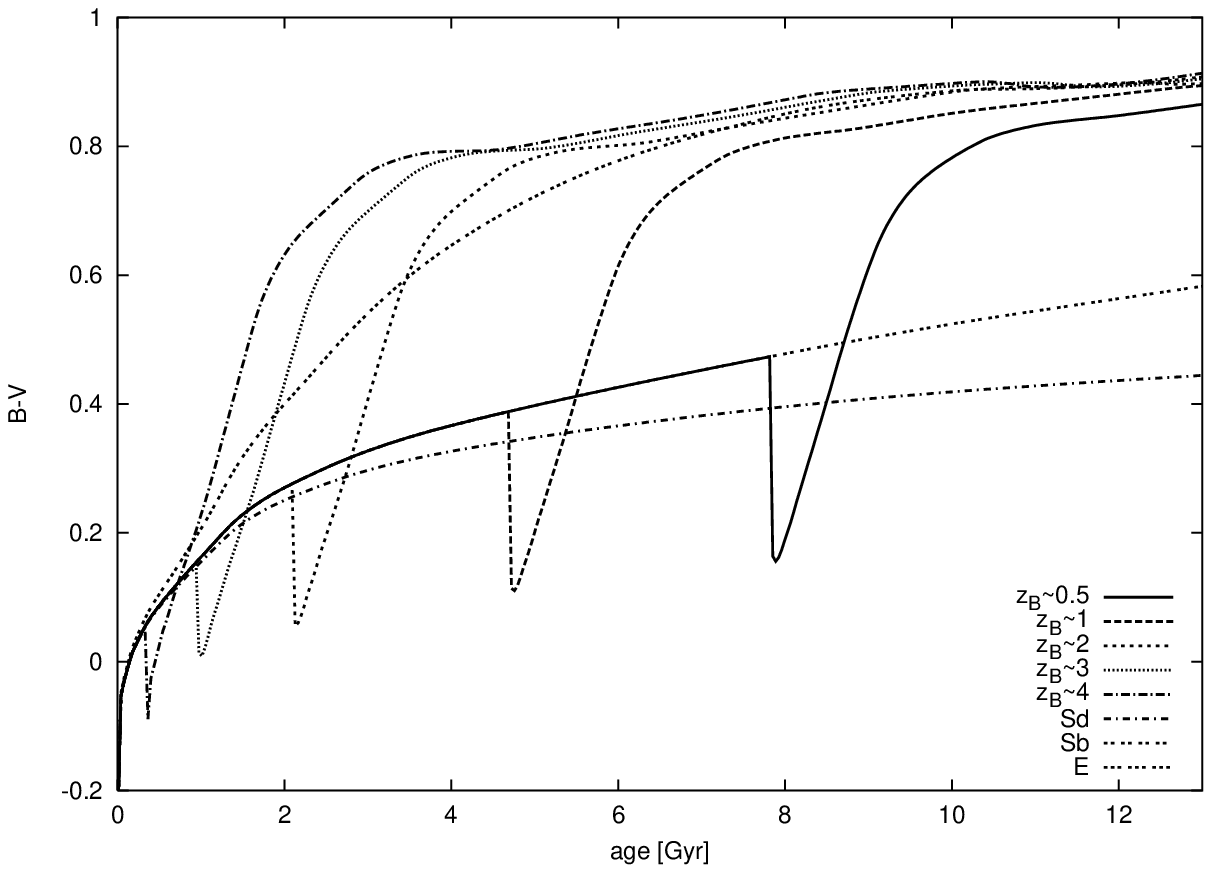,width=8.cm,angle=0}
  \caption{Time evolution of the absolute B-band luminosity ${\rm M_B}$ (top
  panel) and of the restframe ${\rm (B-V)}$ color (bottom panel) of an 
  Sb galaxy with intermediate strength bursts at times corresponding to
  redshifts ${\rm z = 4,~3,~2,~1,~and~0.5}$. For comparison, the undisturbed 
  Sd and E models are also displayed to mark the upper and the lower
  luminosity limits and the red and blue color extremes of our undisturbed
  galaxy models.}
  \label{sbbage}	
\end{figure}

In Fig. \ref{sbbage} we show the B-band luminosity evolution as a 
function of galaxy age on the example of an Sb galaxy with 
intermediate-strength bursts at various times corresponding to redshifts 
${\rm 4 \geq z \geq 0.5}$. 
After an initially rapid increase, the luminosity of the undisturbed Sb galaxy 
reaches a maximum around 2 Gyr. In the following 10 Gyrs its absolute
luminosity ${\rm M_B}$ 
fades by $\sim 1$ mag. During a burst the galaxy gets significantly brighter, 
as expected. Due to the exponential decline of the burst SFR on a timescale of
$2.5 \times 10^8$ yr, the luminosities 
after the bursts fade rapidly below the level of the 
undisturbed Sb galaxy. The bursts in Fig. \ref{sbbage} all have the same 
strength of ${\rm b = 50}$\% in our definition of gas consumption. 
At younger ages, when the galaxy has more gas 
available, the luminosity increase is higher for a given burst strength 
than at older ages, when the gas reservoir is smaller.    

We clearly see that none of the bursts gets as bright as a classical
initial collapse elliptical model (the lowest curve in Fig. 2, upper panel) plotted 
for comparison.
Long enough (${\rm \sim 1/2~t_{Hubble}}$) after early bursts, the Sb postburst
model gets fainter, without any further SF, in ${\rm M_B}$ than even the
faintest and latest-type undisturbed galaxy model Sd. 
We recall that our undisturbed model galaxies are
scaled to match the mean present day absolute B-band luminosities 
${\rm \langle M_B \rangle}$ of the respective 
galaxy types in the local universe.  

The time evolution of the ${\rm (B-V)}$ color shows  
that bursts occurring at later phases, although transforming smaller absolute 
amounts of gas, cause stronger blueing, since by that
time the galaxy itself is already relatively
red than bursts occurring at earlier stages, when the galaxy itself is still
young and blue. 
On the other hand, the reddening after the burst is much stronger during the
long evolution after early bursts than during the shorter time interval between
a more recent burst and the present.  
A color of ${\rm (B-V) \sim 0.9}$ is a red limit to any old, passively evolving 
stellar population. Within a Hubble time galaxies cannot get any redder than
that, even after $\gta 6$ Gyr of passive evolution without any SF.   

It is seen that during and up to $\sim 600~$Myr after the bursts, the Sb+burst
models get significantly bluer than an undisturbed Sd. Shortly thereafter,
however, 
the postburst models reach colors as red -- or even much redder after early
bursts -- than those of a passively evolving elliptical. Note that while the
blueing during a burst is a relatively short-time phenomenon lasting $\lta 300 - 500$
Myr, the red and very
red phases after the end of a burst can last very long, i.e. for many Gyr, 
unless a galaxy resumes
SF from gas (re-)accreted after the end of the burst.

\begin{figure}
  \centering \epsfig{file=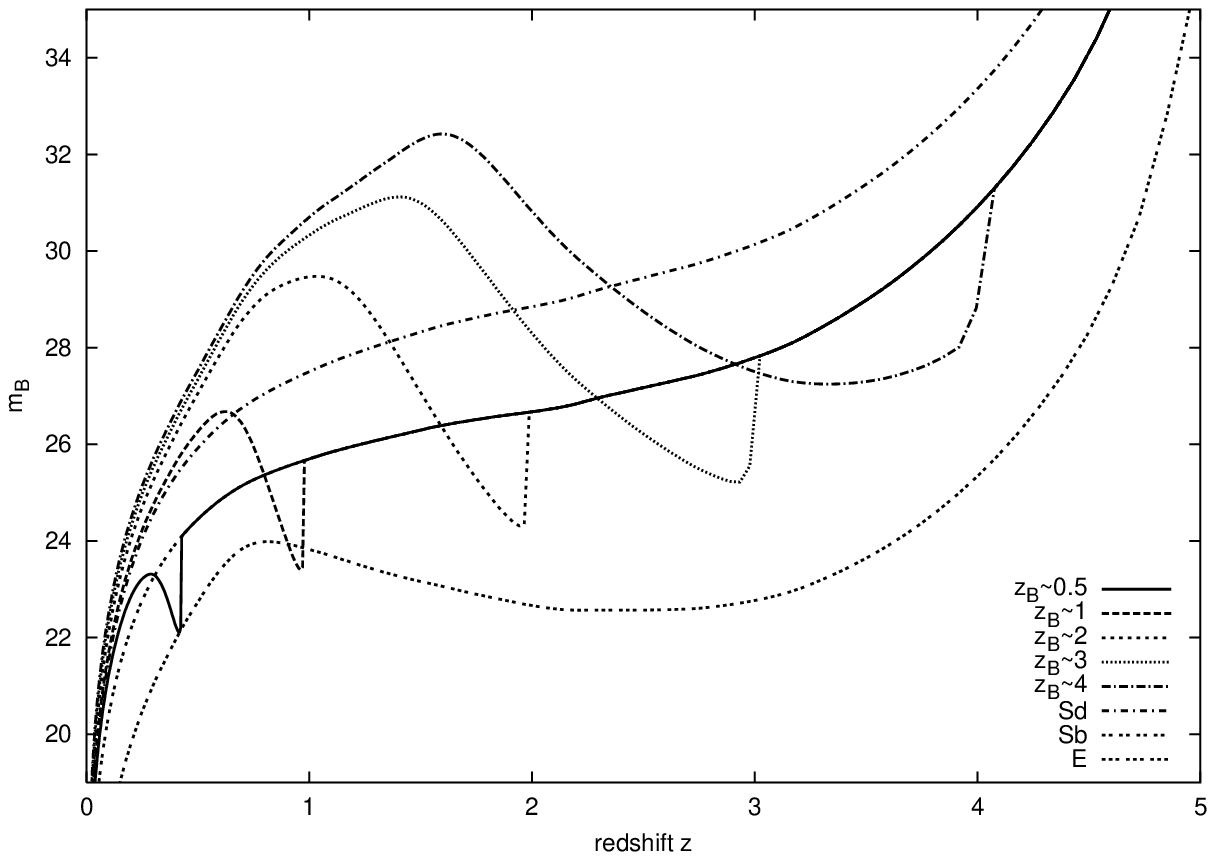,width=8.cm,angle=0}
  \centering \epsfig{file=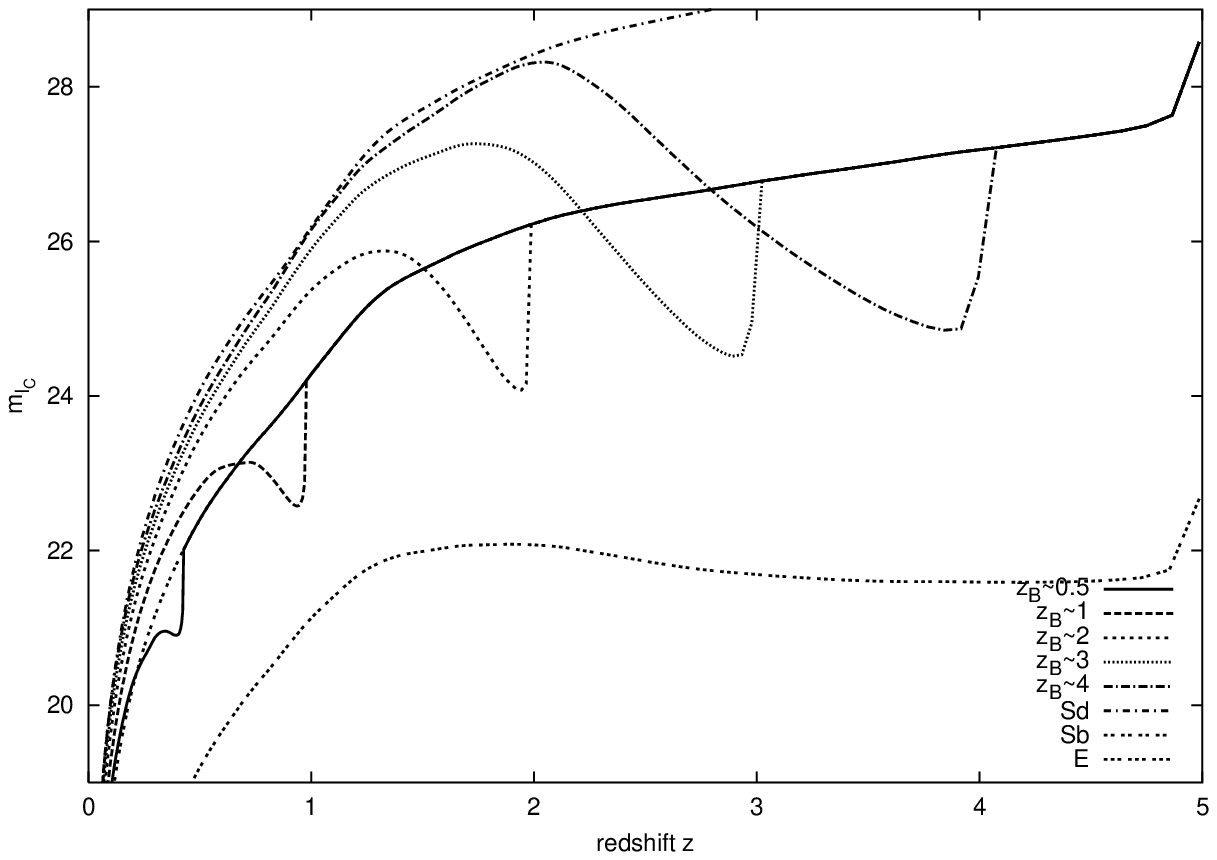,width=8.cm,angle=0}
  \caption{Redshift evolution of an Sb galaxy with intermediate strength bursts 
  at ${\rm z = 4,~3,~2,~1,~and~0.5}$ in terms of observer-frame, apparent B-band
  luminosities ${\rm m_B}$ (upper panel) and in terms of observer-frame, apparent
  ${\rm I_C-}$band luminosities ${\rm m_{I_C}}$ (lower panel). 
  Undisturbed Sd and E models are again displayed for comparison.}
\end{figure}

Figure 3 shows the evolution of the same models as a function of 
redshift. The luminosity evolution, however, is now given in terms of 
observer-frame apparent Johnson B-band and Cousins I-band luminosities (upper
and lower panels, respectively). Now, the luminosity evolution is
no longer only an evolutionary effect, but also includes the cosmological 
bandshifts, dimming, and attenuation. The behavior of the 
bursting galaxies is similar to that explained above. Note, however, the very 
strong apparent fading in ${\rm m_B}$ of galaxies after starbursts at high
redshift by
the combined effects of fading and bandshift; e.g., an Sb galaxy with an 
intermediate strength starburst at ${\rm z \sim 3}$ gets brighter by about 
2.5 mag in ${\rm m_B}$ until a
redshift ${\rm z \sim 2.5}$ and then becomes very faint -- almost 4 mag fainter
in ${\rm m_B}$ than it was before the burst and almost 5 mag fainter by a 
redshift around 1.5 than an undisturbed
Sb galaxy. The same galaxy with
the same starburst around ${\rm z \sim 2}$ would be about 4 mag fainter 
by ${\rm z \sim 1}$ than an undisturbed galaxy. 

Another effect results from the nonlinear transformation between time and 
redshift. While the e-folding time is the same for all bursts 
($2.5 \times 10^8$ yr), 
the duration -- in terms of redshift -- varies. A burst occurring at
${\rm z = 4}$ will light up the galaxy until ${\rm z = 3}$. 
In contrast, a burst at ${\rm z = 1}$ will 
light up the galaxy only up to a redshift ${\rm z=0.8}$. 

As seen in the lower panel of Fig. 3, the effects of starbursts on apparent 
I-band luminosity are qualitatively similar to those in B, but much smaller 
both in terms of brightening during and fading after the burst. 

\begin{figure}
  \centering \epsfig{file=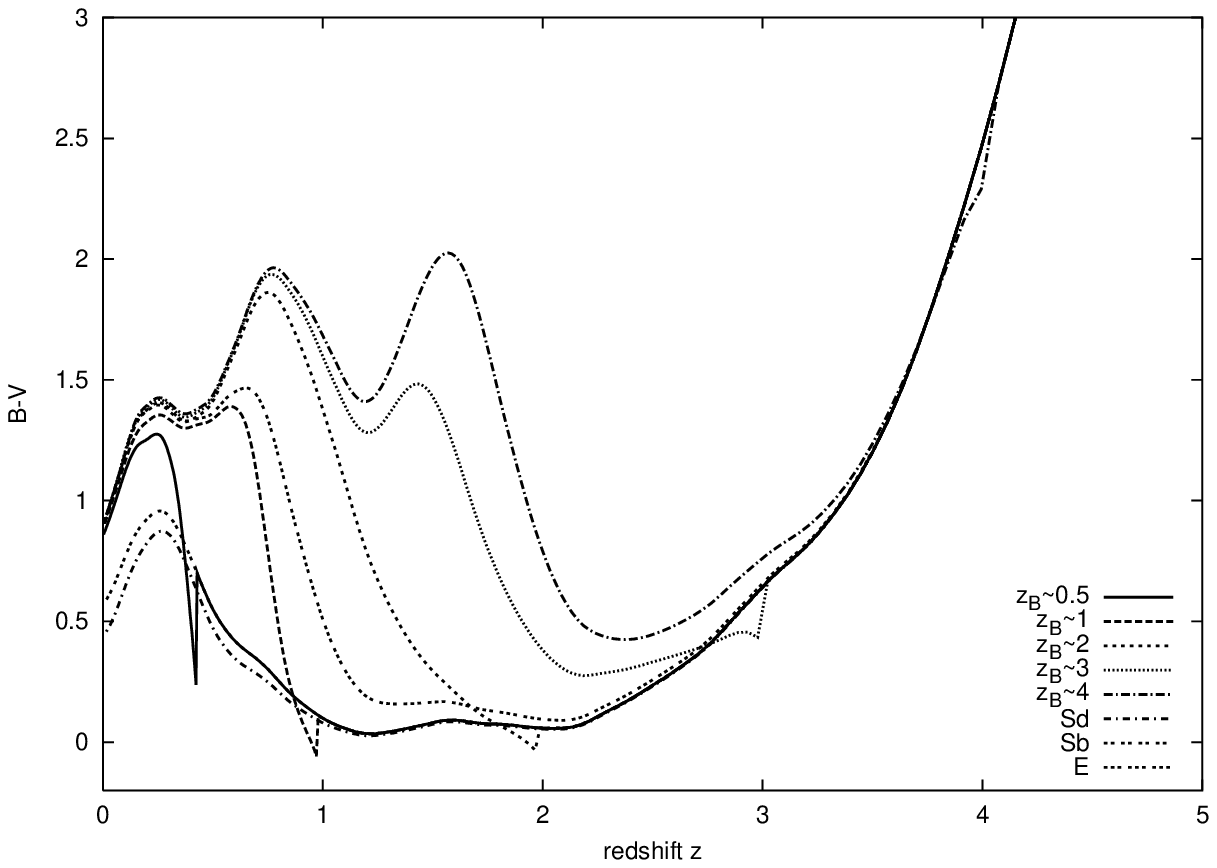,width=8.cm,angle=0}
  \centering \epsfig{file=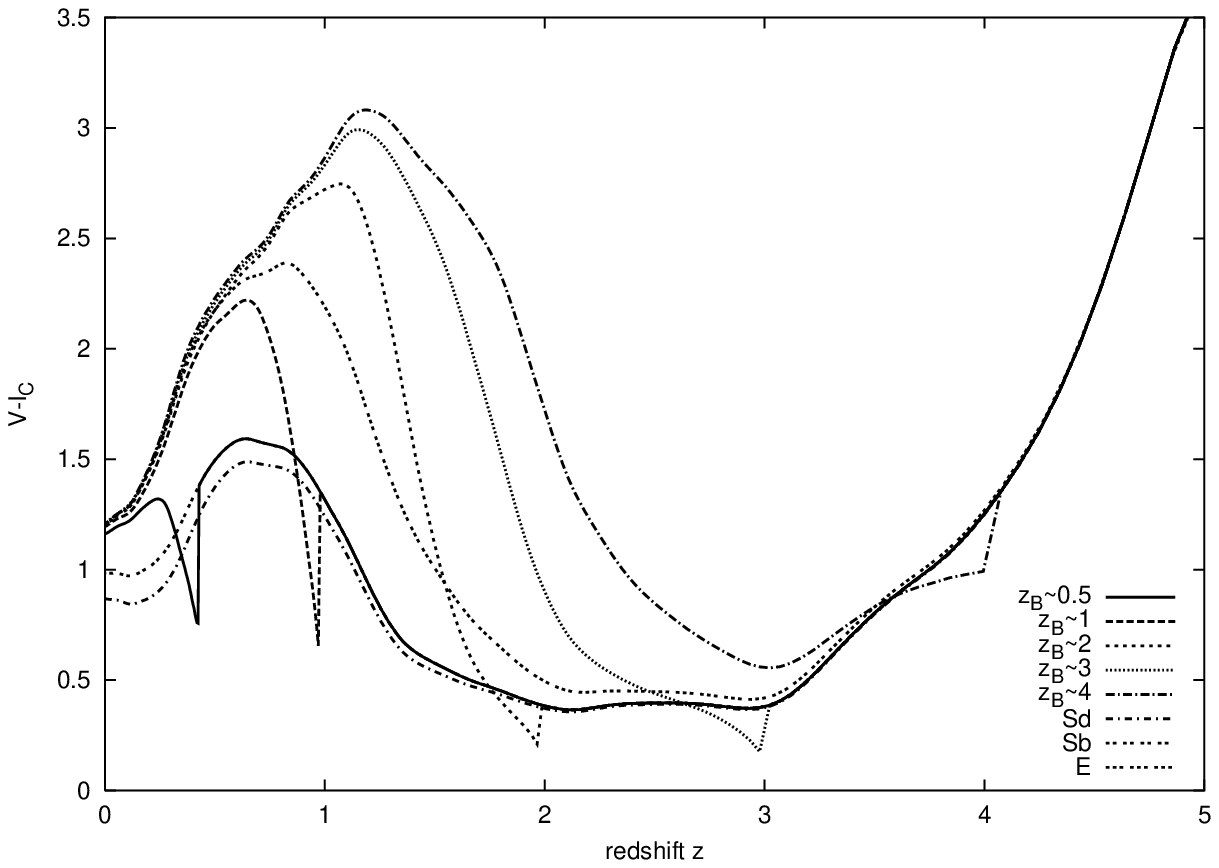,width=8.cm,angle=0}
  \caption{Redshift evolution of Sb models with intermediate-strength
   starbursts at different redshifts in observer-frame ${\rm B-V}$ (upper
   panel) and ${\rm V-I_C}$ (lower panel).}
\end{figure}

Figure 4 shows the redshift evolution of 
${\rm (B-V)}$ and ${\rm (V-I_C)}$ colors in observer-frame. 
This is more complicated to understand than the time evolution in Fig. 2. 
Both bandshift effects and attenuation
play a role. Those Sb galaxies with bursts at redshifts $\gta 3$ get bluer not only
during the active burst but also in the postburst phase until redshifts 
${\rm z=2~-~2.5}$, in line with the undisturbed Sb models. Nevertheless, 
these postbursts are always slightly redder
than the undisturbed galaxy. Bursts at ${\rm z = 4, ~z = 3,~and~z = 2}$ become very red,
up to ${\rm (B-V) \sim 2}$ by redshifts ${\rm z = 2~,z = 1.5,~and~z = 1}$, respectively,  
through ${\rm z \sim 0.5}$. In terms of ${\rm (V-I_C)}$, these same postburst 
galaxies can reach values as red as ${\rm (V-I_C) = 2~-~3.2}$ over the redshift 
range ${\rm 2\gta z\gta 0.5}$.   

In these color vs. redshift plots (Fig. 4), it is clearly
seen that the reddening after starbursts at high redshifts ($4\gta z\gta 0.5$) 
has a very strong effect, one much stronger than the blueing during the burst.
Comparison with the classical initial collapse model for ellipticals also
plotted in these figures shows that, provided SF stops completely, the post-starburst  models
reach colors at redshifts between $0.5\lta z \lta 2$ that are redder than those
of the passively evolving E-model by 0.5 to 2.0 mag in ${\rm (B-V)}$ and by 1 to 1.5 in
${\rm (V-I)}$. The postburst models reach color ranges never attained by
undisturbed or passively evolving galaxies: ${\rm (B-V) \gta 1.5}$ and ${\rm
(V-I_C) \gta 1.5}$ at ${\rm 0.5 \lta z \lta 2}$ for bursts at ${\rm 2\lta z \lta 4}$.
We caution that these red and faint postburst stages will only be reached by
galaxies that do not resume any residual SF after a burst, e.g. from previously
expelled and later re-accreted enriched gas or from newly accreted unenriched
material. The 
earlier a burst occurs, the more time a galaxy has until today to reassemble a
gas reservoir for subsequent SF, as discussed in the context of secondary disk
rebuilding (e.g. Hammer \etal 2005). On the other hand, the time between a
starburst at ${\rm z\sim 4}$ (or 3) and the redshift of maximum postburst
reddening at ${\rm z\sim 2~...~0.5}$ (or ${\rm z\sim1.5~...~0.5}$) is short. 
In the case
of a burst at ${\rm z\sim 4}$, the phase of maximum reddening starts 1.7 Gyr
after the burst 
and holds on for 5.3 Gyr. After intermediate-strength or strong bursts, 
in particular, which most probably lead to a 
strong galactic wind, as observed in many, if not all, high-redshift 
starburst galaxies, the galaxies
will most probably need more time than this before they will eventually 
be able to restart subsequent SF. We thus expect to really observe 
passive post-starburst galaxies
with very red colors ${\rm (B-V) \sim 1-2}$ and ${\rm (V-I_C) \sim 1.5-3}$ in the
redshift range of ${\rm z \sim 2}$ to ${\rm z \sim 0.5}$.  

\begin{figure}
\centering \epsfig{file=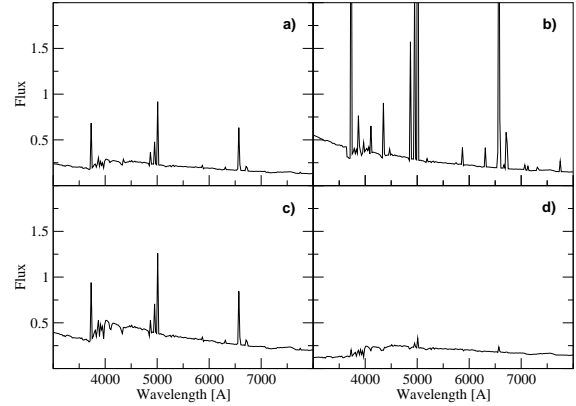,width=9.cm,angle=0}
\caption{Time evolution of the optical spectra for an Sb model with an 
intermediate strength starburst occurring at a redshift ${\rm z=1}$: 
(a) prior to the burst, (b) at maximum burst, (c) 0.5 Gyr after the burst, 
and (d) 1 Gyr after the burst.}
  \label{sbpec}	
\end{figure}

Figure ~\ref{sbpec} briefly highlights the spectral evolution of our models, from
which the photometric properties of redshifted galaxies are derived. The
separate panels show a) the actively SFing Sb galaxy before the onset of the 
burst, b) the
enhanced short-wavelength flux and the strong emission lines during the burst,
c) and d), respectively, the model spectra 0.5 and 1 Gyr after 
the onset of the burst. Panel c) gives a glimpse of the strong higher order
Balmer lines during the post-burst phase and d) shows the almost 
completed transformation from a SFing into a passive K-star-type spectrum with
weak emission lines from the starving star formation still on top. 

\subsection{Effects of the burst strength}
\label{BURSTSTR}
Before, all the bursts we studied had the same strength ${\rm b=0.5}$, i.e.
transforming 50\% of the galaxy's remaining gas reservoir into stars. Now  
we study the influence of 
different burst strengths. Figure 6 shows the redshift evolution during and
after 
bursts with strengths ${\rm b = 0.2,~0.5,~and~0.8}$ occurring in an Sb galaxy 
at a redshift ${\rm z = 1}$. In terms of apparent B-band luminosity, the 
differences are clearly visible, in particular between a weak and an
intermediate strength burst, and less so between the intermediate and strong
bursts (top panel).  

\begin{figure}
\centering \epsfig{file=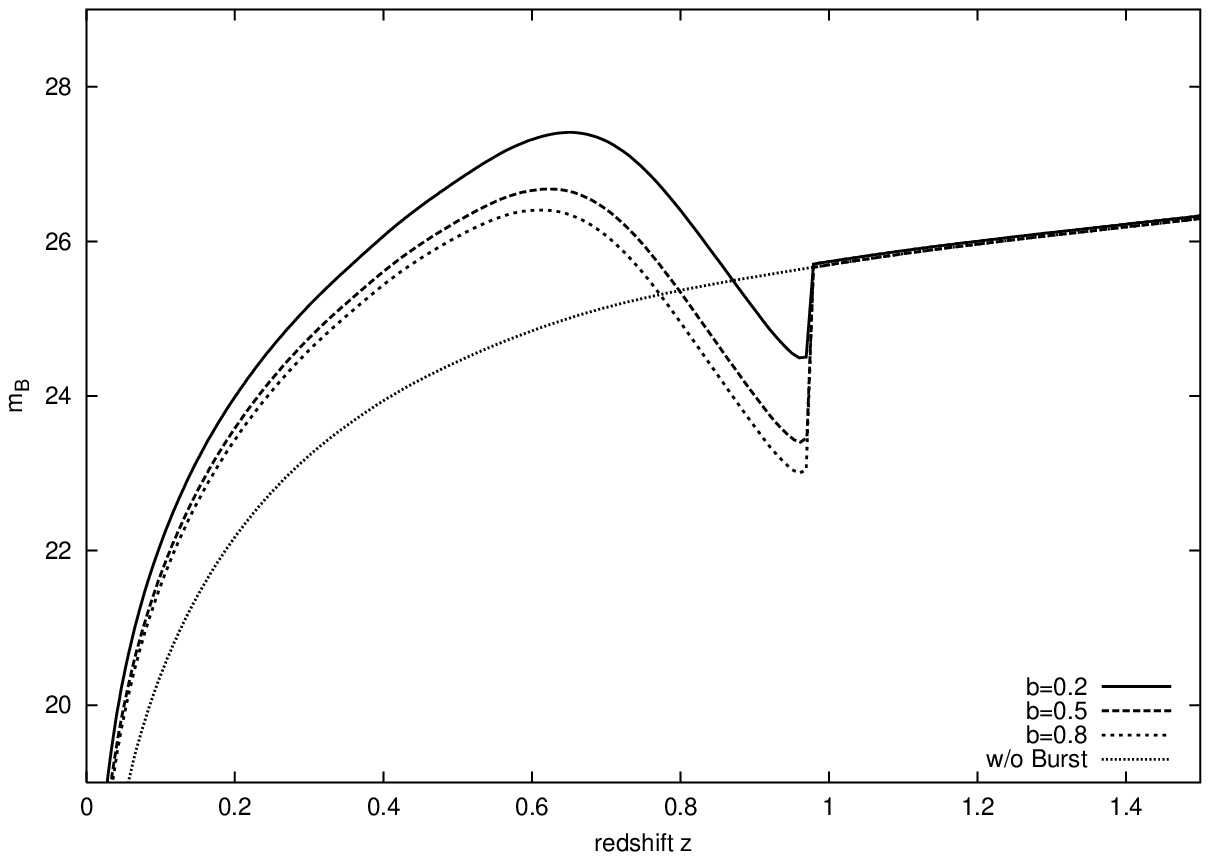,width=8.cm,angle=0}
\centering \epsfig{file=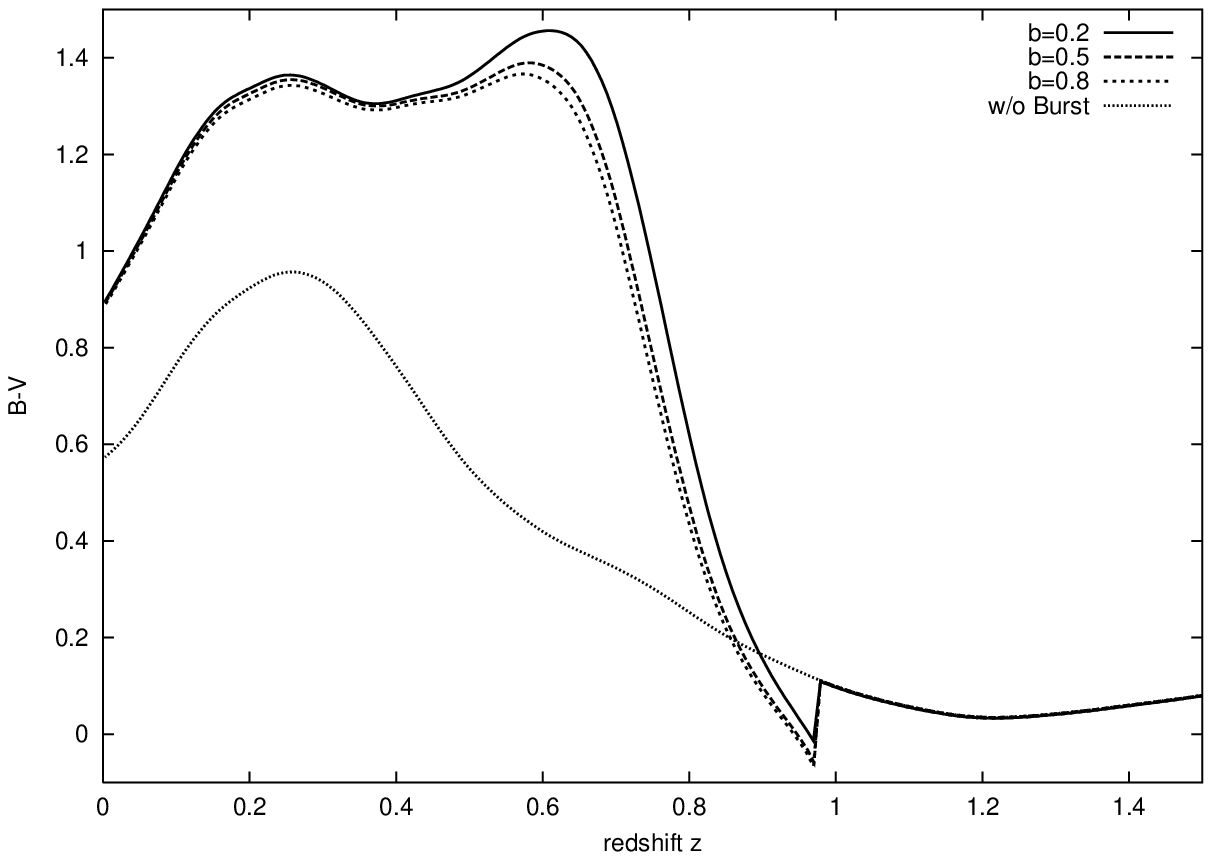,width=8.cm,angle=0}
\caption{The impact of different burst strengths on the redshift evolution of 
$m_b$ (top panel) and on the color evolution ${\rm (B-V)}$ (lower panel) 
for starbursts occurring at ${\rm z=1}$ in an Sb galaxy and 
consuming ${\rm 20\%,~50\%,~and~80\%}$ of the available gas, respectively.}
\end{figure}

In ${\rm (B-V)}$, as in all optical colors, the differences between the 
different burst strengths are fairly small (lower panel). 
Even smaller effects on the colors are observed in our models for bursts occurring at higher
redshifts when the blueing effect is intrinsically smaller because of the bluer
colors of the underlying undisturbed galaxy models.  

\subsection{The role of dust}
\label{DUST}

Although our models at their present stage do not include any dust, we briefly
want to discuss its effects. If we were to include realistic amounts of dust
in our undisturbed models, we would have to slightly adjust the SFRs of the
respective spiral models so as to again reproduce the average
observed colors after a Hubble time (cf. M\"oller \etal 2001). 
Hence, the inclusion of typical, i.e. moderate, amounts of
dust would not affect the evolution of colors by much. During the bursts, dust
could significantly reduce the blueing and the luminosity, in particular in short
wavelength bands, if present in sufficient quantities, as in the
case of dusty starbursts like LIRGs or ULIRGs in the relatively local 
Universe or SCUBA galaxies in the more distant one.
Shortly ($\lta 500$ Myr) after the peak of a starburst, however, local 
example galaxies
appear to have largely or even completely cleared out their dust (cf. NGC 7252,
NGC 3921, . . .). As a result, we expect dust to no longer play any 
significant role during
most of the post-starburst phase, no matter how dusty the starburst phase was. 

\subsection{Relation to EROs}
\label{ERO}
The nature and origin of the Extremely Red Objects, called EROs, defined by
having ${\rm (R-K) \gta 5}$, which are found in ample number densities 
at redshifts
${\rm z\sim 1-2}$, is a topic still under some discussion. Recent results 
suggest that
the ERO galaxy population is composed to approximately comparable parts of 
dusty
starbursts and old passively evolving stellar systems. The passive
galaxies are thought to be the high-luminosity high-mass end of the 
elliptical galaxy
progenitor population at ${\rm 1 \lta z \lta 2}$ and may be related to the
starburst galaxy population at ${\rm z > 2}$ seen in deep submillimeter 
surveys (cf. Mc Carthy 2004 for a recent review). Moustakas \etal (2004)
report $\sim 40$\% of all EROs in the GOODS field to show early-type
morphologies. 

\begin{figure}
\centering \epsfig{file=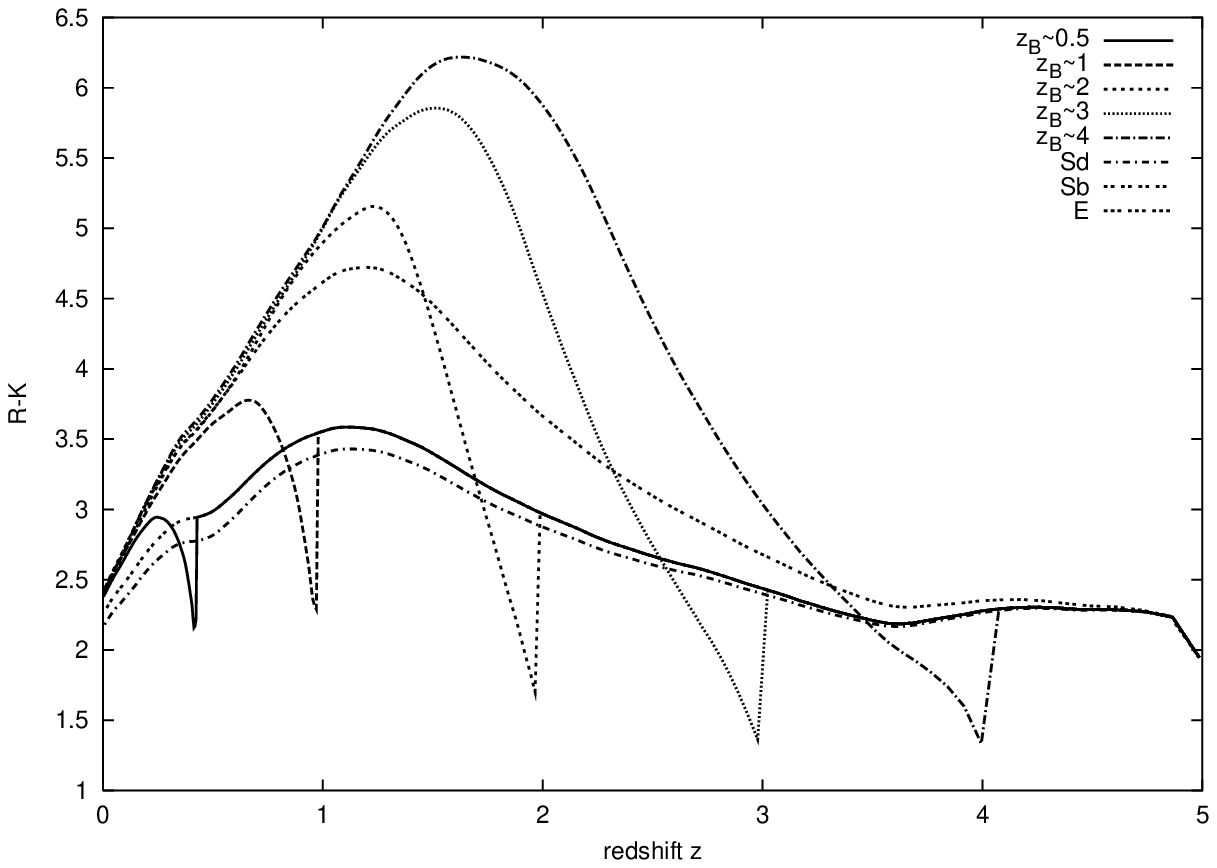,width=8.cm,angle=0}
\centering \epsfig{file=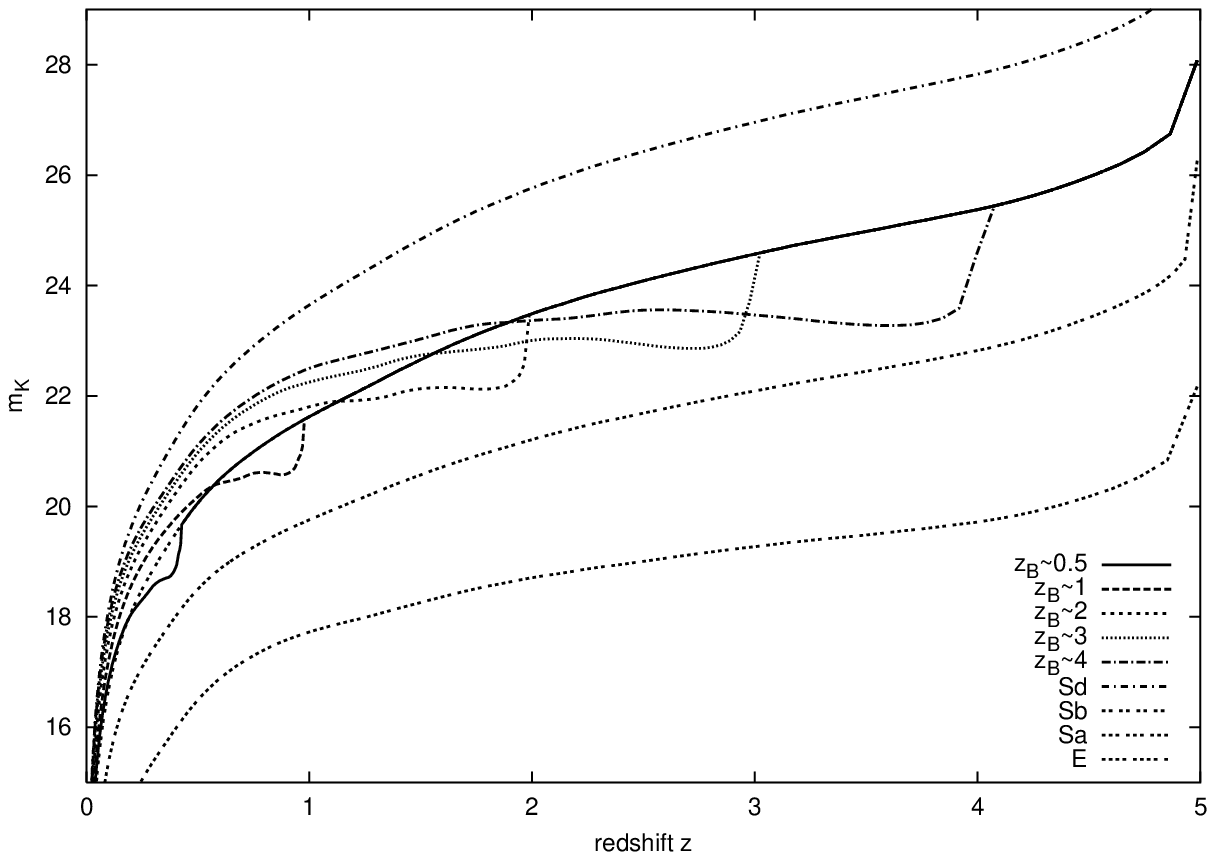,width=8.cm,angle=0}
\caption{Redshift evolution of Sb models with intermediate strength 
starbursts at different redshifts in observer frame ${\rm R-K}$ (top 
panel) and in ${\rm m_K}$.}
\end{figure}

Comparison with our models in terms of ${\rm (R-K)}$ (Fig. 7, upper panel) indicates that
the classical initial collapse model for an elliptical galaxy does not reach 
the color range of EROs. This is only valid, however, for an average luminosity
elliptical. The luminosity-metallicity relation observed for elliptical
galaxies leads us to expect that an elliptical galaxy that by today will be
brighter than average will also have higher metallicity than our model and,
hence, reach redder colors and possibly go though an ERO stage in the course
of its evolution -- provided it formed in this classical scenario and
evolved in isolation, i.e. without accretion of gaseous matter triggering
star formation. Our postburst models with ${\rm 2 \lta z_{burst} \lta 5}$, 
however, all reach far into the ERO color range between ${\rm z\sim
2.3~and~z\sim 1}$. 

Comparison with the classical E-model shows that these very
red colors are only achieved by stellar populations that formed the bulk of
their stars within a short time interval on the order of a few $10^8$ yr and 
rapidly
dropped their SFRs to zero thereafter. The tail of the 1 Gyr exponentially 
declining SFR in our classical E-model is sufficient to keep colors out of 
the extremely red ${\rm (R-K)}$ range forever. 

Another property of observed EROs is their high luminosity ${\rm K
\gta 20~mag}$. In the lower panel of Fig. 7 we plot the redshift evolution 
of apparent
K-band magnitudes ${\rm m_K}$ for our Sb+intermediate strength burst models,  
as well as for our undisturbed E, Sa, and Sd models. 
The K-band (and
other NIR) luminosities increase during bursts, although somewhat
less than e.g. the B-band luminosity. 
An intriguing feature 
of NIR luminosities, in contrast to blue and optical
ones, is that they do not significantly decrease after the burst,
but rather remain fairly constant instead. 
This is due to the well-known fact, that NIR
light traces mass and that mass is dominated by long-lived low-mass stars, 
while blue 
and optical light instead traces the short-lived higher-mass stellar component.
Looking at the K-band evolution of our models in the lower panel of Fig. 7, we 
note that our intermediate strength bursts in an average-luminosity Sb galaxy 
are not luminous enough to reach the K$<20$ domain. However, if a burst, and
in particular a strong one, takes place in a more luminous early type spiral or 
is triggered by a major merger that doubles the luminosity, the K$<20$
criterium of EROs is easily met by galaxies with starbursts at ${\rm z \lta 4}$.    

As discussed in Sect.~\ref{DUST} dust should
not be an issue during most of the post-starburst evolution, so we do 
expect our dust-free models to give approximately correct colors and 
luminosities in this
stage. Even if occurring in a spiral-type galaxy, we expect a reasonably
strong burst -- which, at high redshift, is most probably triggered by 
interactions and mergers during 
hierarchical structure formation -- to be accompanied by a
restructuring of the remnant towards an E/S0 morphology type, in agreement with
the observations by Moustakas \etal (2004) mentioned above.  
While our dust-free models do not give any clue to the dusty-starburst part
of the ERO galaxy population, they show that the $\sim 50$ \% so-called 
passively evolving stellar systems are explained well by our dust-free 
post-starburst models and not by average-luminosity classical E-models with an 
initial collapse and subsequent passive evolution. Our results indicate, that the ${\rm K < 20}$ objects
require strong -- probably merger-induced -- starburst progenitors. 
While brighter-than-average, classical, passively evolving elliptical models
might work due to the color-luminosity relation, we propose here a new
scenario for ERO galaxies, the dust-free post-strong-starbursts, as a third 
alternative to the conventionally assumed, passively evolving vs.
dusty-starburst dichotomy. 

\subsection{First comparison with HDF data}
\label{HDF}
The incentive for the present investigation of the role of starbursts on
the photometric evolution of high redshift galaxies came from a comparison of
our chemically consistent evolutionary synthesis models for undisturbed galaxies
with data for intermediate and high redshift galaxies in the Hubble Deep Field
North (HDF-N) (cf. Bicker \etal 2004). While we found good agreement between our chemically
consistent galaxy models and the bulk of the HDF data over the redshift range
from ${\rm z>4}$ through ${\rm z \gta 0}$, there remained a number of galaxies
 at intermediate and high redshifts with bluer colors than in our bluest undisturbed Sd model described by a
constant SFR and fairly low metallicity, as
well as quite a few galaxies with colors significantly redder in the redshift
range ${\rm 0.5 \lta z \lta 4}$ than our reddest
undisturbed elliptical galaxy model, despite its high metallicity. This lead us
to investigate to what extent starbursts can explain the ´´bluer than
normal¸¸ galaxies and post-starbursts the ´´redder than normal¸¸ ones. 

\begin{figure}
\centering \epsfig{file=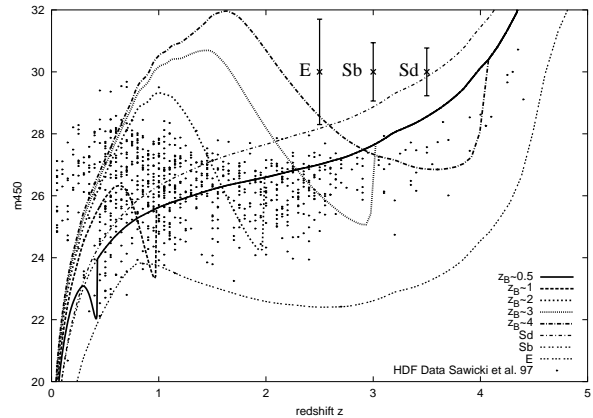,width=8.cm,angle=0}
\caption{Redshift evolution of our models in terms of HST 450W apparent
luminosities $m_{m450}$ compared 
to HDF-N galaxies with photometric redshifts from Sawicky et al. (1997).} 
\label{sbbHDF}	
\end{figure}

Therefor, we now compare our starburst and post-starburst models with photometric data 
for galaxies in the HDF using the photometric redshifts given by Sawicki \etal
(1997). For this comparison, we calculate the 
corresponding photometry from our model spectra explicitly
for the HDF filters in the AB-mag system to avoid 
uncertainties involved in a transformation from HST to Johnson filters. 
In Fig.~\ref{sbbHDF} the observed m450 ($\sim$ B) magnitudes
are plotted together with the redshift evolution of our burst and post-starburst
models and with
our undisturbed E, Sb, and Sd models for comparison. The bars in Fig.~\ref{sbbHDF} 
show the $1\sigma-$luminosity ranges for the respective undisturbed local galaxy
types. We caution that these may change with redshift  
due to bandshift effects.  

The undisturbed model for the classical
elliptical galaxy is much brighter than any observed galaxy at redshifts 
greater than $z=2$. This indicates that the classical monolithic initial 
collapse 
model is not a likely scenario for today's elliptical galaxies. More than half
of the HDF galaxies at ${\rm 0.5 \lta z \lta 1.2}$ are fainter in m450 ($\sim$
B) than our faintest galaxy model Sd. The brighter of these could be
actively SFing dwarfs, while the fainter ones in any case must be passive
galaxies with very little emission in rest-frame U.   
At the lowest redshifts ${\rm z<0.5}$ the selection criteria for the HDF 
result in a
lack of bright galaxies.  
Sb models with intermediate strength bursts easily reach the  brightest 
galaxies in the HDF-N at all redshifts during their active burst phase, as well
as almost all of the faint galaxies at redshifts between ${\rm z = 1.5}$ and
${\rm z = 0.5}$.

\begin{figure}
\centering \epsfig{file=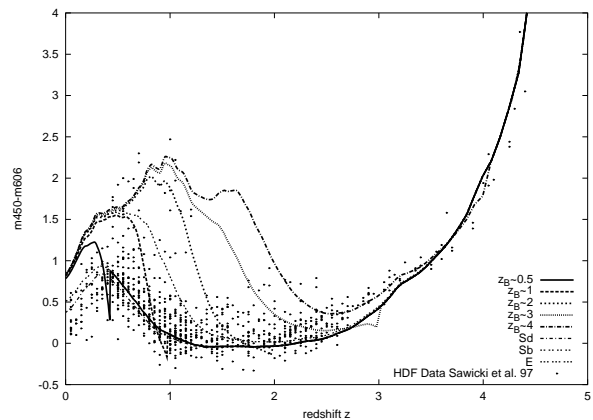,width=8.cm,angle=0}
\caption{Color evolution in terms of (m450$-$m606) for starburst and
undisturbed models compared to HDF-N data.}
\label{sbbvHDF}	
\end{figure}

More information is obtained from a comparison between observations and model
galaxies in terms of colors. In Fig.~\ref{sbbvHDF}, the redshift evolution
of (m450$-$m606) ${\rm \sim (B-V)}$ is plotted. While the bulk of the observed 
galaxies fall well within the color range of our undisturbed galaxy models, 
some are bluer than even our bluest Sd-model over the entire redshift range from
${\rm z > 4}$ through ${\rm z \sim 0}$ and some are significantly 
redder than our undisturbed E-model that marks the red envelope of our 
undisturbed models. Around a redshift of ${\rm z = 1}$ there
are a number of galaxies with ${\rm (m450 - m606) \gta 1.5}$ that 
cannot be reached by
our undisturbed E model, but only by post-starburst galaxies after 
bursts at redshifts ${\rm z \gta 2}$. Essentially all 
galaxies in the redshift range 
${\rm 1.5 \lta z \lta 3.5}$ with colors much redder than those of a classical
passively evolving elliptical model are reached by post-starburst models. 
In the redshift range ${\rm 1 \lta z \lta 2}$, there seems to be an apparent
lack of galaxies with the colors of our postburst models. If the red galaxies
around ${\rm z=0.5 \dots 1}$ are supposed to be postburst galaxies, there 
should also be some postburst galaxies in this area. A look back at the
luminosity evolution (cf. Fig.~\ref{sbbHDF}) solves this apparent problem:  
in this redshift range the postburst galaxies
are too faint and fall below the detection limit for the HDF. All 
HDF galaxies bluer than our bluest undisturbed Sd model are easily reached 
by ongoing starbursts and at low redshift, even a weak burst  
can cause a significant blueing as shown in
Figs. 4 and 6.       

A detailed and quantitative analysis of the HDF galaxy population 
in terms of undisturbed,
starburst, post-starburst models, SFRs, metallicities, masses, etc., using
our own consistent photometric redshift determinations is currently under way and
will be the subject of a
forthcoming paper.

\section{Conclusions and outlook}
\label{CONCL}
The incentive for this present investigation came from a comparison of
our chemically consistent evolutionary synthesis models for undisturbed galaxies
with data for intermediate and high redshift galaxies in the Hubble Deep Field
North (HDF-N) (cf. Bicker \etal 2004). The agreement of our chemically
consistent models, which account for the increasing metallicities of successive
stellar generations, with the data was much better than for any earlier models
solely using solar metallicity input physics. The inclusion of stellar
subpopulations with lower-than-solar metallicities made the chemically
consistent Sd model bluer at increasing redshifts than an equivalent model with
only solar metallicity. The inclusion of stars with 
supersolar metallicity made our chemically consistent classical elliptical 
model redder at intermediate redshifts than the corresponding model for ${\rm
Z_{\odot}}$. While we found good agreement between our chemically
consistent galaxy models and the bulk of the HDF data over the redshift range
from ${\rm z > 4}$ through ${\rm z \gta 0}$, a number of galaxies remained 
with colors bluer than those of our bluest undisturbed Sd model, that is 
described by a
constant SFR and fairly low metallicity at intermediate and high redshifts, as
well as quite a few galaxies with significantly redder colors in the redshift
range ${\rm 0.5 \lta z \lta 4}$ than is our reddest
undisturbed elliptical galaxy model, despite its high metallicity. This caused 
us to investigate to what extent starbursts can possibly explain the bluer-than-normal 
galaxies and to what extent post-starbursts might explain the 
redder-than-normal ones. 

In this paper, we studied the influence of starburst on the photometric evolution
of galaxies at high redshifts. Based on our evolutionary synthesis code GALEV,
we calculated the spectrophotometric evolution of galaxies bursting at redshifts
0.5, 1, 2, 3, and 4. To translate the time evolution of the models into a 
redshift evolution we used a standard cosmological model with ${\rm
H_0=70,~\Omega_m=0.3, ~\Omega_{\Lambda}=0.7}$ and an assumed redshift of galaxy
formation ${\rm z_f \sim 5}$. The star formation histories of undisturbed galaxy
types E through Sd are defined by the requirement that models after a Hubble
time agree with average observed properties of the respective galaxy types in
terms of (i) broad band colors from UV through NIR, (ii) template UV and 
optical spectra, (iii) emission and absorption line strengths, 
(vi) gas content, and (v) metal abundances. 
Starbursts on top of these undisturbed models are described by a sudden
increase in the SFR followed by a decline on an e-folding timescale of
$2.5 \times 10^8$ yr. Burst strengths are defined by the ratio of stellar mass
formed out of the evolving available gas reservoir. In this definition a burst
of given strength produces more stars in a gas-rich galaxy or in a 
galaxy at a higher redshift than in a gas-poor galaxy or in a galaxy at a lower
redshift. 
From the model spectra we calculated magnitudes and colors in the Johnson-Cousins
and HST filter systems. Neither dust nor AGN contributions have been included in our
models at the present stage. 

Following our models in their time evolution through starburst and 
post-starburst phases, we found the following results:

\begin{itemize}
\item The
blueing and brightening during a burst is a relatively short-term phenomenon 
lasting $\lta 300 - 500$ Myr. The reddening and dimming after the burst are much
stronger, and the red and very
red phases after the end of a burst can last very long, i.e. for many Gyr, 
unless a galaxy resumes
SF from gas (re-)accreted after the end of the burst.

\item Bursts occurring at later phases, although transforming smaller absolute 
amounts of gas into stars, cause stronger blueing, since the galaxy itself is 
already relatively red, than bursts occurring at earlier stages, when 
the galaxy itself is still young and blue. 
On the other hand, the reddening after the burst is much stronger during the
long evolution after early bursts than during the shorter time interval between
a more recent burst and the present.  
\end{itemize}

Putting starburst and post-starburst evolution into a cosmological context
and including the non-linear transformation between time and redshift, evolutionary
effects, cosmological dimming and band shifts, as well as attenuation by
intergalactic neutral hydrogen, we observe the following:

\begin{itemize}
\item A burst of given strength starting 
at ${\rm z = 4,~3,~2,~1,~0.5}$ makes a galaxy brighter in apparent 
${\rm m_B,~m_I}$ and bluer than the undisturbed model until redshifts 
${\rm z \sim 2.8,~2.2,~1.5,~0.7,~0.3}$, respectively. 

\item Bursts at high redshift, although producing larger amounts of stars,  
cause a very small blueing only and do so only in very small redshift intervals.

\item During their post-burst evolution, galaxies with bursts at high redshift 
get very red and remain so until low redshifts.
\end{itemize}

We discussed our results in relation to the Extremely Red
Objects (EROs) characterized by ${\rm (R~-~K) \gta 5}$ and found in ample
numbers at redshifts ${\rm 0.5 \lta z \lta 2}$. We find that:
\begin{itemize}
\item After bursts at 
redshifts between 2 and 4$-$5, our model galaxies, even without dust, 
naturally evolve into the color range of EROs in
the course of their passive postburst evolution. 
The classical initial collapse and passive evolution elliptical model 
does not reach colors as red as those of the EROs, at least not for the 
progenitors of present-day average luminosity ellipticals. Particularly massive
and luminous, and therefor particularly red, monolithic initial collapse 
ellipticals might reach ERO colors in the
appropriate redshift range. They would then, however, be much more luminous at
redshifts $\geq 1$ than any observed galaxy in the HDF. 
The ${\rm K<20}$ EROs require strong bursts on top of early-type spirals 
or the intrinsically brighter merger-induced starbursts as precursers. 
This fits well with the observation that $\sim 40$ \% of all EROs 
show early-type morphologies, if we
bear in mind that a major merger remnant will soon be transformed 
into an early-type galaxy in terms of morphology, as
well as in terms of photometry, as our models show. 
\end{itemize}

The dust-free models
presented here cannot, of course, describe the dusty-starburst fraction 
of ERO galaxies. 

A very first comparison of our models to HDF-N 
galaxies with photometric redshifts from Sawicki \etal (1997) shows that 
almost all of the galaxies that were not explained by our chemically
consistent models for undisturbed galaxies E \dots Sd are naturally
explained by our starburst and post-starburst models: 

\begin{itemize}
\item HDF galaxies bluer than our bluest low-metallicity, 
undisturbed Sd galaxy model, most of which are 
at low redshifts (${\rm z\lta 1}$) and a few also at higher
redshifts, are explained well by ongoing starbursts. 

\item HDF galaxies redder than our high-metallicity, undisturbed, classical elliptical
model, most of which are at redshifts around ${\rm z \sim 1}$, are well-matched 
by post-starburst models. The reddest galaxies around ${\rm z \sim 1}$ 
require models bursting at ${\rm z \gta 2}$.   
\end{itemize}

Dust is not yet included in a self-consistent way in our models at the present
stage. We have argued that the good agreement between the dust-free models and the
data indicates that galaxies detected in UBVI are not strongly affected by dust.
In dusty
starbursts our models would, of course, severely underestimate the true burst 
strength. During post-starburst stages, observations of local galaxies show
that dust no longer plays an important role. Very shortly after a burst,
galaxies apparently  manage to get rid of most of their dust. 

The consistent inclusion of appropriate amounts of dust as well as a more
extensive and quantitative interpretation of intermediate and high-redshift 
galaxies in terms of undisturbed, starburst, postburst galaxies, SFRs,
metallicity and mass evolution, etc., on the basis of our own photometric
redshifts is in progress and 
will be the subject of a forthcoming paper.


\acknowledgement
We thank our anonymous referee for a helpful report that significantly 
improved the presentation of our results and acknowledge the language editor for
numerous suggestions. 
Jens Bicker gratefully acknowledges partial financial support from the 
DFG (Fr 916/10-1-2).

\end{document}